# Contribution to the Formal Specification and Verification of a Multi-Agent Robotic System

**Nadeem Akhtar**
*IRISA – University of South Brittany - FRANCE*
*Assistant Professor the Department of Computer Science & IT*
*The Islamia University of Bahawalpur Bahawalpur, 63100, PAKISTAN*
E-mail: nadeem.akhtar@iub.edu.pk

**Malik M. Saad Missen**
*Assistant Professor, The Department of Computer Science & IT*
*The Islamia University of Bahawalpur Bahawalpur, 63100, PAKISTAN*
E-mail: saad.missen@gmail.com

**Abstract**

It is important to have multi-agent robotic system specifications that ensure correctness properties of safety and liveness. As these systems have concurrency, and often have dynamic environment, the formal specification and verification of these systems along with step-wise refinement from abstract to concrete concepts play a major role in system correctness. Formal verification is used for exhaustive investigation of the system space thus ensuring that undetected failures in the behavior are excluded. We construct the system incrementally from subcomponents, based on software architecture. The challenge is to develop a safe multi-agent robotic system, more specifically to ensure the correctness properties of safety and liveness. Formal specifications based on model-checking are flexible, have a concrete syntax, and play vital role in correctness of a multi-agent robotic system. To formally verify safety and liveness of such systems is important because they have high concurrency and in most of the cases have dynamic environment. We have considered a case-study of a multi-agent robotic system for the transport of stock between storehouses to exemplify our formal approach. Our proposed development approach allows for formal verification during specification definition. The development process has been classified in to four major phases of requirement specifications, verification specifications, architecture specifications and implementation.

**Keywords:** Formal methods, Correctness properties, Safety property, Liveness property, Formal verification, Multi-agent robotic system, Formal architecture, Finite State Process (FSP), Labelled Transition System (LTS), Architecture Description Language (ADL).

## 1. Introduction

Today multi-agent robotic systems are not safe. Human lives can be lost due to errors in these systems, therefore it is important to have multi-agent robotic systems that are safe. Here by safe the emphasis is on correctness properties on the behavior of multi-agent robotic systems, the correctness properties that can be described by a combination of safety and liveness. How can safety and liveness properties be reinforced during the analysis, design, and implementation of a multi-agent robotic system? These properties can be satisfied by having a multi-agent robotic system development approach based on formal methods and languages, having major phases of requirement specifications, verification specifications, architecture specifications, and implementation [Akhtar, 2010].

Our area of research is formal methods for the specification and verification of a multi-agent robotic system. Model-checking has a degree of formalization that gives the flexibility to apply formal methods according to our implementation requirements. Our research domain focuses on formal methods, multi-agent systems, and robotics as shown in fig.1.

An agent is considered as a computer system situated in some environment, capable of autonomous actions in this environment in order to meet its design objectives [Wooldridge and Jennings, 1995]. Multiple agents are necessary to solve a problem, especially when the problem involves distributed data, knowledge, or control. A multi-agent system is a collection of several interacting agents in which each agent has incomplete information or capabilities for solving the problem [Jennings, Sycara and Wooldridge, 1998]. These are complex systems and their specifications involve many levels of abstractions. They have concurrency, and often have dynamic environments. A multi-agent robotic system is distributed. A distributed system along with the interactions between its components presents a high-level of complexity, and results into complex possible system behavior in different scenarios. Complete understanding of the system behavior is required for the analysis, design, and implementation of such a system. To overcome the complexity problems in them and get significant results with formal analysis, we must cope with complexity at every stage of development: from the specification phase to the analysis, design and verification phase. The formal specification and verification of a multi-agent robotic system along with its step-wise refinement from abstract to concrete concepts plays an important role in system correctness. Safety and liveness properties have to be enforced during each development phase of requirement specifications, verification specifications, architecture specifications, and implementation.

**Fig 1:** Research domain

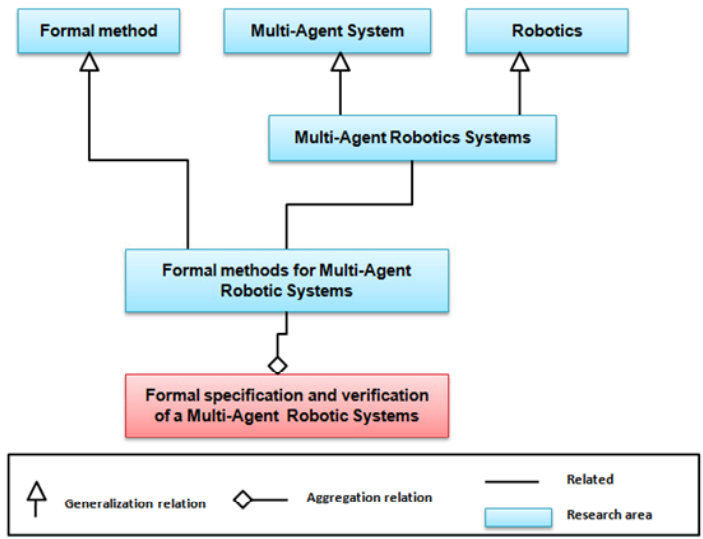

One of the most challenging tasks in software specification engineering for multi-agent robotic systems is to ensure correctness properties of safety and liveness, especially as these systems have high concurrency and in most cases have dynamic environment. Finite automata based model-checking of safety and liveness properties play major role in system correctness i.e. to verify that the code matches its requirement and design specifications is important.

Our system consists of small robotic agents that work in a closed environment. Labelled Transition System (LTS) [Magee and Kramer, 2006] specifications based on Finite State Process (FSP) language have been used for specification definition of our multi-agent robotic system. These automata-based specifications are flexible, rigorous, executable and practical and play vital role in ensuring correctness properties. Therefore by using this automata-based model-checking approach we are able to obtain a concurrent system in which there are processes working in parallel and there are synchronizations between different processes. The LTS and its associated analysis tool LTSA have an incremental and interactive approach to the development of component based systems. Consequently, components can be designed and debugged before composing them into larger systems.

## 2. The Problem Statement, Objectives, and Contributions

Our problem statement is; How can a safe multi-agent robotic system be developed? Here by safe the focus is on correctness properties which can be described by a combination of safety and liveness. Thus the core question is how can safety and liveness properties be enforced during the development of a multi-agent robotic system?

The most challenging task in software specifications definition for robotic multi-agent systems is to ensure correctness. Safety and liveness properties are critical for system correctness. As these systems have concurrency, often have dynamic environments, the formal specification and verification of these systems; the step-wise refinement from abstract to concrete concepts play an important role in system correctness. It is important to address the following issues:

1. The formal specification of our multi-agent robotic system which has a dynamic architecture i.e. which can change during run-time;
2. To support the property-preserving transformations of agents from abstract to concrete specifications to code generation by stepwise refinement;
3. To support system verification using formal model-checking approaches; and to formally check the safety and liveness properties of the system.

In order to address the above issues, a formal approach is required which does not rely solely on immediate software development, but on continuous engineering, adaptation, and evolution of the software system.

Our objective is to propose a development approach that provides formal verification of safety and liveness properties, architecture description, and a service-oriented simulation based system implementation. It results into the development of a multi-agent robotic system that satisfies correctness properties of safety and liveness. Another objective is the formal specification, architecture, and implementation by considering the functional properties; by refining in stepwise phases from abstract to concrete specifications along with the formal verification of these specifications.

This approach supports the efficient formal requirement specifications, verification specifications, architecture specifications, transformations, refinement from abstract to concrete concepts, and implementation of the system. The work aims to define and develop a formal architecture-based approach for the engineering of a multi-agent robotic system. The formal verification specifications i.e. verifying correctness properties of safety and liveness have been defined by labelled transition system based on finite state processes. For Formal architecture specifications, p-ADL dot NET [Oquendo, 2004] based formal architecture is specified. The system is implemented by Service-Oriented Architecture (SOA) based simulation.

Our contributions are;
1. An approach based on a combination of methods to allow for formal verification and evaluation during development phases of requirement specifications, verification specifications, architecture specifications, and implementation;
2. Checking correctness properties of safety and liveness at each development phase;
3. A multi-agent robotic system case study to exemplify each phase of this approach;

4. A combination of process algebra and finite automata based techniques to define the formal specifications of our system and verifying each flow of concurrent executions.

## 3. Background Studies
### 3.1. Formal Methods

Formal methods are based on a solid mathematical foundation. Formal specification has a precise mathematical semantics which in turn support formal verification. Formal verification allows mathematical rigorous proofs that specifications are according to the objectives, code is according to the specification, and code produces only the results that are required. These methods can achieve complete exhaustive coverage of the system thus ensuring that undetected failures in behavior are excluded. The core objective of a solid formal approach is to provide unambiguous and precise specification [George and Vaughn, 2003]. The requirements model based on mathematics create precise specification of the software, and ensure correctness. The formal representation of software requirements provides a way for logical reasoning about the construct produced and this achieves precise description and allows a stronger design that satisfies the required properties. As formal specification and verification techniques are getting more accomplished and mature, our capabilities to design and develop complex systems are also maturing and growing quickly. Formal notations are used to produce a complete detailed representation of the system that helps in the understanding, design, and development of the system. The requirements for distributed, large, and complex systems are complicated, problematic at the initial stages and evolve periodically throughout the life cycle. This creates a need for the method of requirement implementation to be flexible and robust, so that it can easily accommodate the continuous versions of change [Luqi and Goguen, 1997].

To overcome the complexity problems in multi-agent systems and get significant results with formal specifications, we must cope with complexity at each phase: requirement specification phase, architecture specification phase to design and implementation phases. We must assure formal verification during all phases. Formal verification can achieve complete exhaustive coverage of the system thus ensuring that undetected failures in the behavior are excluded. We can prove the correctness of agent software systems by formalizing critical components in the multi-agent development life-cycle. The reasons to have formal software engineering methods are:

- Rigorous analysis of system properties;
- Property-preserving transformations and error-free implementation
- High quality of each phase of the development process;
- Firm foundation for the adaptation and evolution process;
- Continuous correctness especially as multi-agent robotic systems are concurrent and often have dynamic environments;
- Formal specification and modeling of a multi-agent system architecture which can change at run-time;
- Specification according to the functional and non-functional properties;
- Property-preserving step by step transformations from abstract to concrete concepts, then stepwise refinement to implementation code;
- Improved documentation and understanding of specifications.

Model-checking [Berard et al., 2001] [Clarke, Grumberg, and Peled, 2000] is a type of formal method used to verify concurrency properties; it can be viewed as exhaustive investigation of a system state space to prove certain correctness properties. Process calculi based symbolic techniques such as π–ADL [Oquendo, 2004], CSP [Hoare, 1978], CCS [Milner, 1980], ACP [Bergstra and Klop, 1987], and LOTOS [Van Eijk et al., 1989] provide formal specifications for complex systems. Here complex means a system with a large number of independent interacting components, with concurrency between components and constant evolution.

### 3.2. Correctness Properties: Safety and Liveness

The safety property is an invariant which asserts that "something bad never happens", which means that an acceptable degree of system working state is maintained. [Magee and Kramer, 2006] have defined safety property S = {a1, a2, … , an} as a deterministic process that asserts that any trace having actions in the alphabet of S, is accepted by S. ERROR conditions are like exceptions which state what is not required, as in complex systems we specify safety properties by directly stating what is required. The liveness property asserts that "something good happens" which describes the states of a system that an agent can bring under certain conditions. Progress property P = {a1, a2, … , an} defines a property P which asserts that in an infinite execution of the system, at least one of the actions a1, a2, … , an will be often executed infinitely [Giannakopoulou, Magee, and Kramer, 1999]. These properties play a vital role in system verification. Safety and liveness properties are complementary to each other, and both together are vital to ensure system correctness.

### 3.3. Gaia Multi-Agent Method

The Gaia [Zambonelli, Jennings, and Wooldridge, 2003] requirement specifications recognize the organizational structure as the core concept for the development of an agent system. A suitable choice of this organizational structure is required to meet the functional requirements. It is based on organizational abstractions to drive the analysis and design of a multi-agent system, and it considers a multi-agent system as a computational organization consisting of interacting roles. These organizational abstractions play a significant role in the analysis, design, and implementation of a multi-agent system in a complex environment. For Gaia, the word method is used instead of methodology, as we consider the term methodology to be used for the study of methods. It has a concrete syntax, which can be extended to deal with the formal specification aspects of a multi-agent system, and it generates a number of models and specifications that can be used by different software development methods for implementation. After the completion of the design phase, we have a well-defined collection of agent roles to implement, and can define the agent and service model. During the specification definition we move from abstract to concrete concepts, these abstract concepts conceptualize the system while the concrete concepts are used during the design phase, and are related to implementation.

   The result of the design phase could be easily implemented in a technology neutral way. It captures the organizational structure of the system which allows for going systematically from the requirement analysis to a comprehensive design. For precise and unambiguous specifications, we need to formalize the specifications of each component and process.

**Table 1:**   Gaia: Abstract and concrete concepts [Zambonelli, Jennings, and Wooldridge

| **Abstract concepts** | **Concrete concepts** |
|---|---|
| Roles | |
| Permissions | |
| Responsibilities | Agent types |
| Liveness properties | ServicesAcquaintances |
| Protocols | |
| Activities | |
| Safety properties | |

   a. A successful correct method has a well-defined formal basis. Formal basis provides the precise understanding of the terms and concepts used in a method [Wooldridge and Jennings, 1995].
   b. Once we have designed the specification, we must be able to implement a correct system with respect to this specification. The next issue is to propose an approach to move from an abstract specification towards a concrete model. Manually refine the specification into an executable form via some formal refinement process.

1. Directly execute the abstract specification along with its animation
2. Translate the specification into a concrete model using automatic translation techniques.

c. It can play a significant role in the analysis and design of dynamic and open systems. In these systems components can join and leave the environment at runtime, and are composed of sub-components that may be different at design time and run time. They are a complicated class of systems to engineer [Gasser, 1991] [Hewitt, 1991].
d. The organization structure is implicitly defined in the role and interaction models. These structures capture and represent the organization's communication and control structures.

## 3.4. Labelled Transition System (LTS)

Labelled transition systems [Magee and Kramer, 2006] are mathematical objects for the formal verification and evaluation of concurrent systems. It is founded on model-checking for the verification of concurrency properties; it represents the system as a set of interacting finite state machines along with their properties; it exhaustively explores the system state space to prove correctness properties of safety and liveness, and it performs compositional reachability analysis to exhaustively search for violations of these properties. [Magee and Kramer, 2006] proposed an analysis tool LTSA [LTSA, 2006] shown in fig.2, that generates labelled transition system consisting of parallel composition of asynchronous processes, interleaving interaction-shared actions.

**Fig 2:** The toolkit LTSA

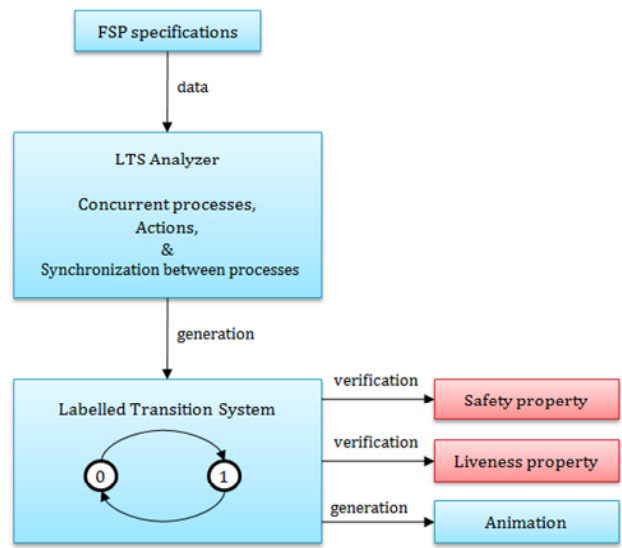

As a result we are able to obtain a concurrent system in which there are processes working in parallel and there are synchronizations between different processes. LTSA also provides specification animation for an interactive exploration of system states.

FSP is a process algebra notation having finite state processes used for the concise description of component behavior particularly for concurrent systems. It is a finite-automata based method that provides construct to formalize specifications of software components and architecture. Each component consists of processes; each process has a finite number of states and is composed of one or more actions. There exists concurrency between elementary calculator activities for which there is a need to manage the interactions, communication and synchronization between processes.

## 3.5. π-ADL

The π-ADL [Oquendo, 2004] provides the software engineer with the fundamental structure and behaviors constructions for describing static as well as dynamic software architectures. It is an executable formal specification language and supports automated analysis as well as refinement of dynamic architectures. The π-ADL has as mathematical foundation the higher-order typed π-calculus [Sangiorgi, 1992] [Milner, Parrow and Walker, 1992]. It is a well-formed higher-order calculus for defining dynamic and mobile architectural elements, which takes its foundation in work related to the use of π-calculus as a semantic foundation for architecture description languages [Chaudet and Oquendo, 2000] [Chaudet et al., 2000]. According to [Milner, 1999], a natural solution for specifying dynamic behavior is π-calculus as it provides a computation model which is Turing-complete. It is an ideal choice for describing concurrent processes that communicate through message passing. In π-calculus every computation can take place but it is not always easy to demonstrate and express. π-ADL is a language having both structural and behavioral architecture-centric constructs, defined as an extensive version of the higher-order typed π-calculus. Fig-3 shows the architectural concepts of π-ADL.

**Fig 3:** Architectural concepts in π-ADL [Oquendo, 2004]

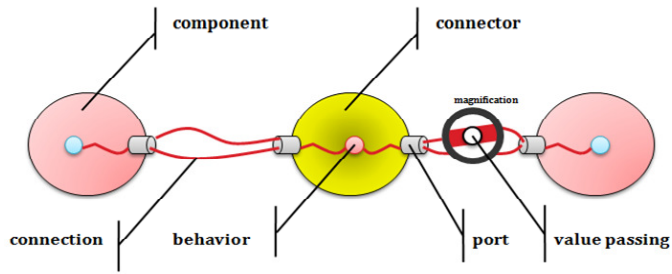

It achieves high architecture expressiveness and Turing completeness with the help of a simple formal syntax notation. As with any design of a language, the design of π-ADL makes tradeoffs between competing requirements and constituencies:
1. Making the language well suited for machine-automated processing for enactment, analysis, refinement and evolution vs. as a stand-alone language for humans: π-ADL is specified in a layered-approach with a core canonical abstract syntax and formal semantics, and different concrete human-oriented notations [Oquendo, 2005].
2. Making the language well suited for software architects to design large-scale software vs. making it automatic semantically tractable: π-ADL is based on a compositional approach [Oquendo, 2005].

According to [Oquendo, 2005] the π-ADL design follows the following language design principles [Morrison, 1979] [Sangiorgi, 1992] [Strachey, 1967] [Tennent, 1977]:
1. *Correspondence principle*: the uses of names are consistent in π-ADL. Particularly there is a one to one relationship between the method of introducing names in declarations and parameter lists;
2. *Abstraction principle*: all major syntactic structures have abstractions defined over them e.g. π-ADL supports abstractions over behaviors as well as abstractions over data;
3. *Data-type completeness principle*: each data-type is a first-class citizen having no restrictions on its use.

## 4. The Proposed Approach

An approach has been proposed for the formal specification and verification of multi-agent robotic system. The requirements are specified, formally verified on the basis of safety and liveness properties, the architecture is specified, and the system is implemented. Our proposed approach is a combination of multi-agent methods, languages, and techniques; which takes into account the safety and liveness properties at each phase of development. This approach is exemplified by a case study of a multi-agent robotic system [Akhtar, 2010].

Our approach starts by the identification of components and sub-components of the system i.e. each and every part of the system that can be formally defined. Each component is formally verified and validated, particularly the critical components. The approach consists of four main development phases of requirement specification, requirement verification, architecture specification, and system implementation as shown in fig.4.

It is exemplified by a case study which is a multi-agent system composed of robotic agents. The mission is to transport goods from one storehouse to another. The robotic agents named as carrier agents transport goods from one storehouse to another. There is a possibility of collision between these carrier agents and collision resolution techniques are applied to avoid system deadlock.

**Fig 4:** A detailed view of the proposed approach

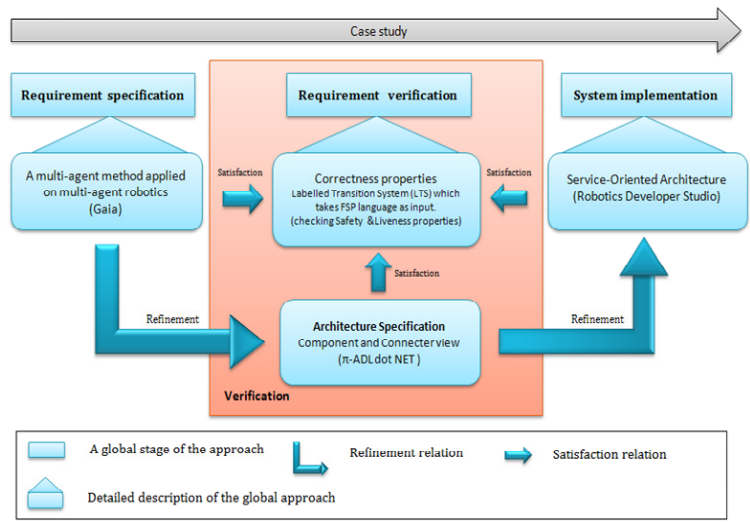

The requirements are specified by using Gaia [Zambonelli, Jennings, and Wooldridge, 2003]. Gaia has a concrete syntax to express properties, and it is suitable to model behaviors. Formal verification of correctness properties of safety and liveness is done by defining the system as a labelled transition system which uses FSP as input. FSP based on process algebra is a formal language which is specifically useful for specifying concurrent behavior. LTS has processes executing concurrently, with each process having one or more actions and synchronization between parallel processes by action sharing. During verification each sub-portion of the system is formally verified to make it consistent with the rest of the system, and at the end the system is verified as a whole. Transformations are made from the Gaia requirement specifications to LTS verification specifications for formal verification of the system. The architecture is specified by π-ADL dot NET language which provides a formal executable architecture model consisting of abstractions and behaviors. These architecture specifications describe the static, as well as dynamic aspects of architecture. The system is implemented by service-oriented based C# simulation implementation.

### 4.1. Requirement Specification

The requirement specification phase starts with the identification of early requirements. It is followed by the specification of a multi-agent system as an organization. In this organization, there are multiple abstraction levels. The organizational rules are defined, which put forth the global system properties; global relationships between roles; global relationships between protocols; global relationships between roles and protocols; and constraints within which the system has to work. The environmental model, which studies the environment and entities related to it, is defined. The role model has responsibilities and permissions, the responsibilities are expressed by safety and liveness properties. Agent roles are also defined. A single agent can have one or more roles but a single role cannot be performed by more than one agent. Safety and liveness properties are defined in this initial phase along with the definition of agent roles. At this phase these properties can be defined by regular expressions or by first-order predicate logic. Protocols are defined between agent roles which define the interactions between agents. A services model is defined, where each service is defined according to input, output, pre-condition, and post-condition.

**Fig 5:** Requirement specification phase

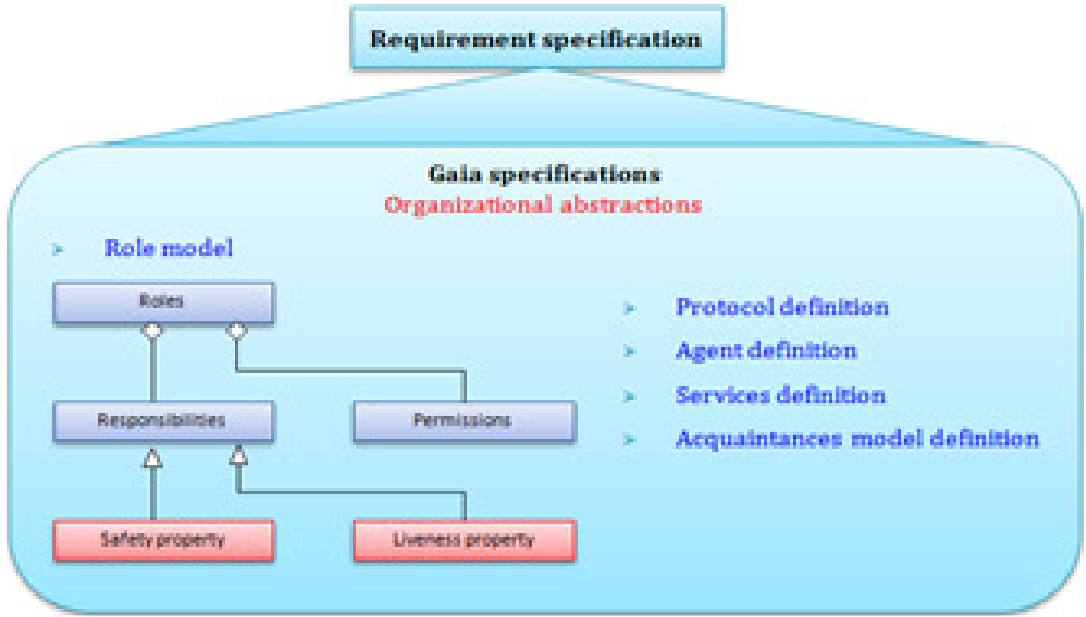

These requirement specifications have a well-defined formal semantics. They capture the organizational structure of the system. They allow for going systematically from the requirement analysis to a comprehensive design. For precise and unambiguous specifications, we need precise mathematical semantics. Organization structures are defined implicitly inside these requirement specifications, within the role and interaction models. It is important to have a precise knowledge of the terms and concepts of a method. Once we specify the system, we would be able to implement a system that is correct with respect to our specifications. The next step is to move from abstract specifications to a concrete computational model.

During requirement specification phase, the safety properties are defined using first-order predicate logic, while liveness properties are defined using regular expressions. Here, it is to be noted that the Gaia role model does not have constructs for the formal verification of safety and liveness properties; therefore the formal verification of these properties is carried out in the next phase of our approach. The agent model identifies the agent instances; and at the end, the acquaintances model is defined which gives a global picture of agents, their environment along with their interactions. The major emphasis throughout this phase is on the safety and liveness properties.

### 4.2. Requirement Verification

Major emphasis is put on the requirement verification and the safety and liveness properties defined in the requirement specification phase are verified in this phase. The system is broken down into sub-components. Each component is verified by a formal model-checking exhaustive method. This involves exhaustive verification of all the states, processes, and actions of each component along with its sub-component. After that all the sub-components are assembled together and the system is verified as a whole [Akhtar, Guyadec, and Oquendo, 2009].

Finite state process is a process algebra notation used for the concise description of component behavior particularly for the concurrent systems. It has strong artifacts for construction of concurrent processes, and therefore it is ideal for concurrent systems. It provides the constructs to formalize the specification of software components, each component consists of processes and each process has a finite number of states and is composed of one or more actions. The processes are modelled as a sequence of actions, and formal specification of dynamic behavioral aspects of the multi-agent robotic system are provided, correctness properties of safety and liveness are verified, along with progress property, deadlock freedom, and sequencing constraints. Concurrency exists between elementary calculator activities; processes are sequential or concurrent and there is management of the interactions, communication, and synchronization between processes.

The correctness properties of safety and liveness defined in requirement specification phase along with the multi-agent system environment are now specified as a labelled transition system. There are actions, processes, states, and transitions between states. The verification specifications developed are a discrete system with a trace of actions; there are parallel processes with synchronization between them by action sharing.

**Fig 6:** Requirement verification phase

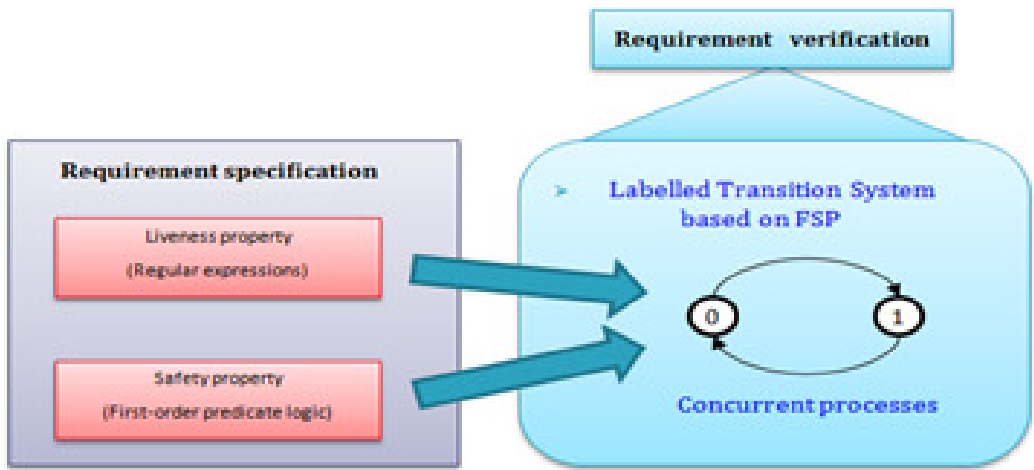

### 4.2.1. Moving from Requirement Specification to Requirement Verification

There is a satisfaction relation between requirement specification and requirement verification. This satisfaction relation is the formal verification of the two correctness properties of safety and liveness. This satisfaction relation is exemplified by a case study. The Gaia role model liveness and safety properties along with the organizational rules are specified in the form of finite automates for verification.

**Fig 7:** Moving from requirement specification to requirement verification

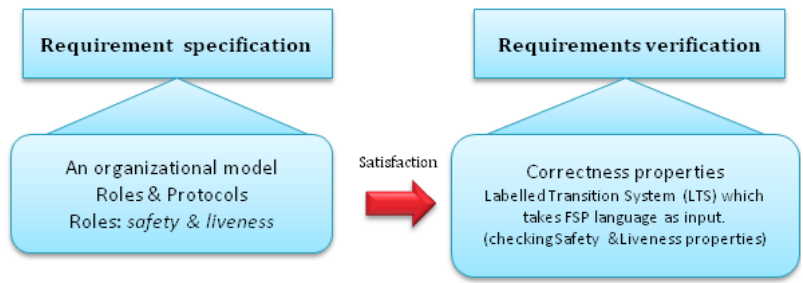

**4.3. Architecture Specification**

A formal architecture [Akhtar, Guyadec, and Oquendo, 2012] has been proposed which specifies the static, as well as the dynamic aspects of the system. In this architecture, the architectural elements are identified. All these architectural elements are separately specified and then connected together to represent the system as one unit. The system architecture is based on π-ADL dot NET which is a dot NET extension of π-ADL [Oquendo, 2004]. We have an architecture consisting of abstractions and behaviors that is formal, consisting of components and connectors, that executes and that can change dynamically during the executions.

1. These architecture specifications provide a formal system having a mathematical foundation that can be used to describe static as well as the dynamic software architecture.
2. They have as formal foundation the higher-order typed π-calculus [Sangiorgi, 1992]. It is a well-formed higher-order calculus for defining communicating and mobile architectural elements.
3. They focus on the formal specification of architecture from the run-time perspective: the run-time structure, the run-time behavior, and the evolution of architecture over time.
4. They are executable i.e. a virtual machine executes the software architectures specifications.
5. They support multiple concrete syntaxes: both textual and graphical notations.
6. They support automated verification of properties by model checking.

**4.3.1. Moving from Requirement Specification to Architecture Verification**
The architecture specifications are based on requirement specifications. When we move from requirement to architecture then the safety and liveness properties should be preserved. The whole system is represented in the form of π-ADL dot NET with emphasis on safety and liveness properties.

**4.3.2. Moving from Architecture Specifications to Simulation Implementation**
The system is implemented as a simulation which reflects the architecture specifications; from π-ADL based system to Service-Oriented Architecture (SOA) based robotic simulation system. There should be conformations between the properties of architecture specification and simulation implementation.

**4.4. System Implementation**

The system is simulated based on SOA with each component as a service, with components having one or more sub-components with every sub-component implemented by a service. These services are loosely integrated and are orchestrated together by an orchestration service. As a result, we have a system that has reusable components.

This services based simulation is implemented by programming C# based Microsoft Robotics Developer Studio (MRDS) services. A refinement relation has been defined between the architecture specifications and these C# based services. It is an implementation of the LTS specifications in a simulation environment. In our system, each and every application is a service. An application is a composition of loosely-coupled concurrently executing components. For example: the carrier robot consists of a number of services orchestrated together. It has two wheels with a motor, sensors comprising of two bumpers for collision detection, and infrared laser for distance measurement and collision avoidance. Each of these motor, bumper, and infrared lasers is implemented by a service. There is a service for the orchestration of these sensors, motor, and actuator.

**4.4.1. Moving from Implementation to Verification Specification**
The robotics simulation implementation specifications must satisfy the finite automata based LTS system. Both implementation and verification specifications should preserve the safety and liveness properties. These LTS properties should also be preserved during the simulation implementation.

The simulation is continuous with a continuous flow of actions. Each part of the simulation is a service along with the orchestration of services. On the other hand, the LTS based system is a much lower abstraction level; has concurrent processes; each process having discrete actions. There are discrete states and the system moves from one state to another. The simulation which is continuous system must satisfy the discrete LTS system.

**Fig 8:** Satisfaction relation from implementation to requirement verification

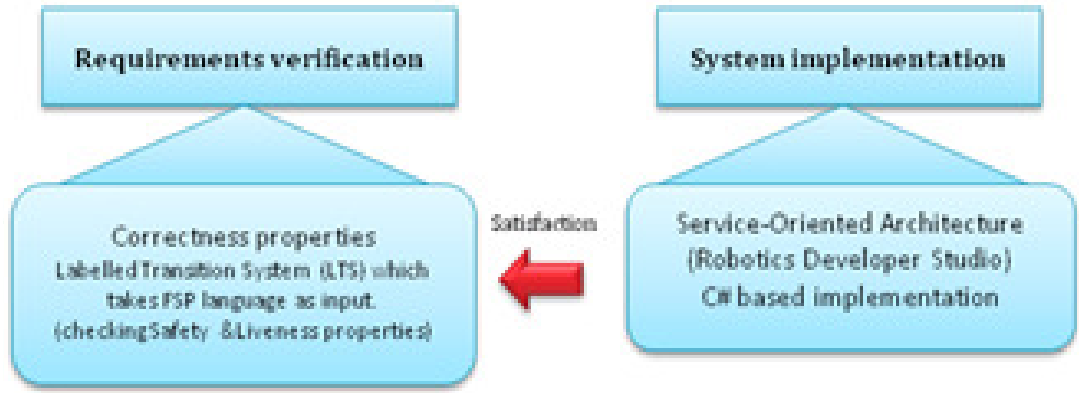

The safety and liveness properties should be preserved in both, and there should be a clear relationship between the two systems. The simulation specifications are able to satisfy the verification specifications. In our simulation model, we create a trace of actions that is equivalent to the trace of actions created by LTS specifications. A mapping of activities provides trace equivalence among requirement specification, verification specification, architecture specification, and implementations.

## 5. Case Study: Multi-Agent Robotics Transport System

In this section we present a case study of multi-agent robotics system. It is a system composed of robotic transporting agents. The objective is to specify our system and then verify the correctness properties of safety and liveness. The mission is to transport stock from one storehouse to another. They move in their environment which in this case is static i.e. topology of the system does not evolve at run time. There is a possibility of collision between agents during the transportation. Collision resolution techniques are applied to avoid system deadlocks. We have specified each and every part of the system i.e. agents along with the environment in the form of LTS.

### 5.1. Types of Agents

There are three types of agents
1. *Carrier agent*: It transports stock from one store-house to another; can be loaded or unloaded and; can move both forward and backward direction. Each road section is marked by a sign number and the carrier agent can read this number.
2. *Loader / Un-loader agent*: It receives/delivers stock from the storehouse, can detect if a carrier is waiting (for loading or unloading) by reading the presence sensor, it ensures that the carrier waiting to be loaded is loaded and the carrier waiting to be unloaded is unloaded.
3. *Store-manager agent*: manages the stock count in the storehouse and it also transports the stock between storehouse and loader/un-loader.

## 5.2. Environment

There is a road between storehouse-A and storehouse-B which is composed of a sequence of interconnected sections of fixed length. Each road section has a numbered sign, which is readable by carrier agents. There are three types of road sections depending upon the topology of the road as shown in fig.9. Each of the three types of road sections has a unique numbered sign. The road is single lane and there is a possibility of collision between agents. There is a roundabout at storehouse-A and storehouse-B.

**Fig 9:** Environment consisting of road partitions

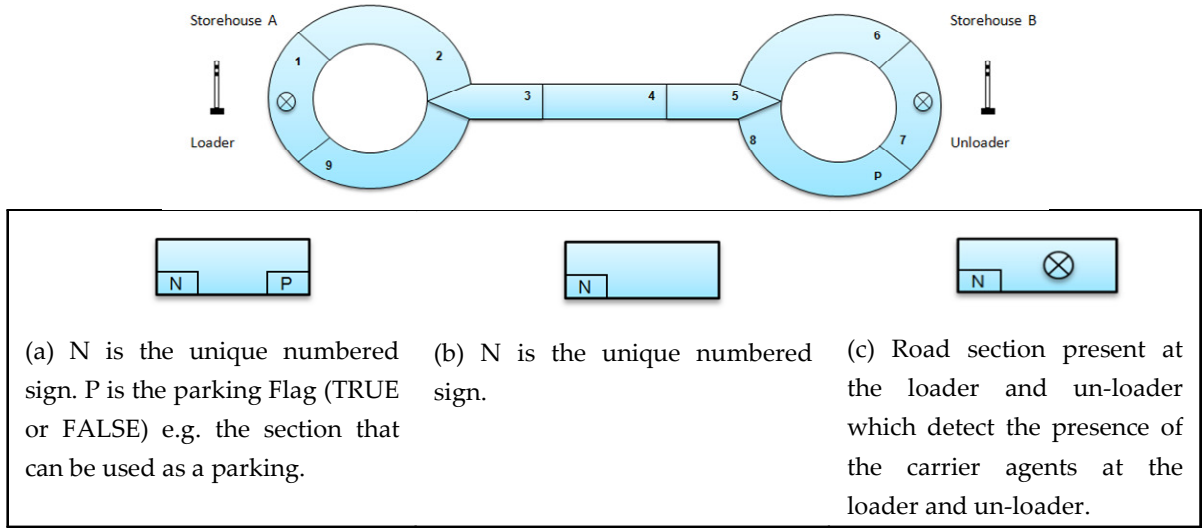

(a) N is the unique numbered sign. P is the parking Flag (TRUE or FALSE) e.g. the section that can be used as a parking.

(b) N is the unique numbered sign.

(c) Road section present at the loader and un-loader which detect the presence of the carrier agents at the loader and un-loader.

## 5.3. Scenario

In this case study we have used a road topology consisting of nine road partitions to represent all states and processes as shown in fig.9. It is the smallest circuit (i.e. combination of road partitions) that allows us to study all properties that would be in a much larger circuit. We have considered the case in which initially storehouse-A is full and storehouse-B is empty. The carrier task is to transport stock from storehouse-A to storehouse-B until the storehouse A is empty. Loader at the storehouse-A loads, and the un-loader at the store-house-B unloads the carrier agent. The store-manager keeps a count of stock in each storehouse. In this case the environment is static. At the central section (3, 4, 5) there is a possibility of collision between carrier agents coming from the opposite directions. Priority is given to the loaded carriers i.e. if there is a collision between a loaded and an empty carrier than the empty carrier moves back and waits at the parking region during which the loaded carrier passes and unloads. The parking region as shown in the fig.9 consists of the road partition 8.

## 6. Gaia Based Requirement Specifications

The major part of the work is to take the Gaia specifications and then use them in a way that they can be verified by using FSP language. Gaia method as described in section-4 consists of a number of models, we may be looking into only the roles model and interaction model which constitutes the analysis phase of Gaia.

## 6.1. Agent Roles

The role of an agent defines what it is expected to do in the organization, both in concert with other agents and in respect of the organization itself. Often an agent's role is simply defined in terms of the specific task that it has to accomplish in the context of the overall organization. Organizational role models precisely describe all the roles that constitute the computational organization. They do this in terms of their functionalities, activities, responsibilities as well as in terms of their interaction protocols and patterns. In the role model the liveness and safety expressions play important role for system verification.

In our system for the carrier agent there are move_full and move_empty roles. These roles are better adapted for this type of route where priority is given to the loaded carriers. Here in this paper due to space constraints we present the Move_full role of our system i.e. role of a loaded carrier agent moving from Storehouse-A to Storehouse-B.

**Table 2:** Move_full

| Role Schema: Move_full |
|---|
| Description:<br>Role of a loaded carrier moving from storehouse A to storehouse B. |
| Protocols and Activities:<br>readSign, movetoNext, collisionSensorTrue, carrierWait, readUnloadSign, waitforUnloading, unloadCarrier |
| Permissions:<br>reads:<br>sign_number (external)<br>collision_sensor (internal)<br>changes:<br>position (internal)<br>next position (external) /// (True or False) checks if the next position is available |
| Responsibilities:<br>Liveness:<br>Move_full = Move.(readUnloadSign.waitForUnloading.unloadCarrier)<br>Move = (readSign. movetoNext)+<br>\| (collisionSensorTrue.Wait).(readSign.movetoNext)+<br>Wait = carrierWait+<br>Safety:<br>(sign number $\in \{2,\ldots,6\} \Rightarrow$ isLoaded)<br>   (sign number $\in \{2,\ldots,6\} \wedge$ next position = sign number+1 $\Rightarrow$ isLoaded) |

Here activities (underlined) are ReadSign, MovetoNext, CollisionSensorTrue, CarrierWait, and ReadUnloadSign. And there are two protocols WaitforUnloading and UnloadCarrier WaitforUnloading: when a loaded carrier reads the unload sign i.e. it reaches the unload road partition, it stops there and waits until it is unloaded.

Consider the Liveness property of the Move_full role. It shows all the activities and protocols that make up the role. The carrier has two choices, first it can read sign and move to the next road partition, second it detects the collision sensor then it waits, at the end it reads the unload sign i.e. at the road partition in front of the un-loader, and in this case the carrier stops and waits for being unloaded, so now it's no more a loaded carrier and is no more part of the Move_full role.

The safety property is an invariant which states that any carrier playing that role schema is currently loaded. Here next_position identifies the direction of the loaded carrier at the route.

## 6.2. Interaction Model

There are dependencies and relationships between the various roles in a multi-agent organization which are the set of protocol definitions, one for each type of inter-role interaction. Here table-1 shows the protocol definitions related to Move_full and Move_empty role.

**Table:**     Move_full role protocols

| waitForUnloading | | |
|---|---|---|
| Move_full | Unload | *sign_number* |
| The full carrier agent waits for the un-loader agent | | *position* |

| Unloading | | |
|---|---|---|
| Move_full | Unload | *sign_number* |
| The full carrier agent waits for the un-loader agent | | *position* |

**Table:**     Move_empty role protocols

| waitforLoading | | |
|---|---|---|
| Move_empty | load | *sign_number* |
| The empty carrier agent waits for the loader agent | | *position* |

| loadCarrier | | |
|---|---|---|
| Move_empty | load | *sign_number* |
| The empty carrier agent is loaded by the loader agent | | *position* |

# 7. LTS Verification
## 7.1. Road – System Environment

In our case study the road is environment and each carrier has its particular route. The route is the path taken by carrier agents on the road to transfer stock from one storehouse to another. The route has been classified in two types the FULL_ROUTE path taken by loaded carriers and the EMPTY_ROUTE path taken by the empty carriers. The carrier agents move on the route in a clockwise direction. Here below are the FSP specifications for the route.

**Fig 10:** LTS specifications of the route (environment)

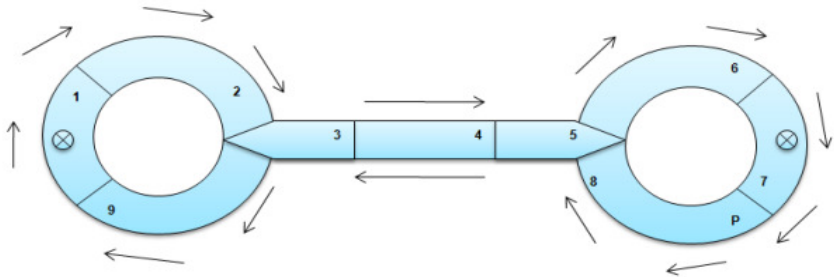

| 1 | range R = 1..9 |
|---|---|
| 2 | **ROUTE = EMPTY_ROUTE**[9], |
| 3 | **FULL_ROUTE**[v:R]=( |
| 4 | when (v==7) readunloadSign -> **FULL_ROUTE**[v] |
| 5 | \| when (v!=7)   readSign[v] -> **FULL_ROUTE**[v] |
| 6 | \| when (v>=1&v<=6)movetonext -> **FULL_ROUTE**[v+1] |
| 7 | \| when (v==7)  waitforunloading -> **EMPTY_ROUTE**[7] |
| 8 | ), |
| 9 | **EMPTY_ROUTE**[v:R]=( |
| 10 | when (v==1)  readloadSign -> **EMPTY_ROUTE**[1] |
| 11 | \| when (v!=1)  readSign[v] -> **EMPTY_ROUTE**[v] |
| 12 | \| when (v==7)  movetonext -> **EMPTY_ROUTE**[v+1] |
| 13 | \| when (v==8)  movetonext -> **EMPTY_ROUTE**[5] |
| 14 | \| when (v==5)  movetonext -> **EMPTY_ROUTE**[v-1] |
| 15 | \| when (v==4)  movetonext -> **EMPTY_ROUTE**[v-1] |
| 16 | \| when (v==3)  movetonext -> **EMPTY_ROUTE**[9] |
| 17 | \| when (v==9)  movetonext -> **EMPTY_ROUTE**[1] |
| 18 | \| when (v==3)  movetoprevious -> **EMPTY_ROUTE**[v+1] |
| 19 | \| when (v==4)  movetoprevious -> **EMPTY_ROUTE**[v+1] |
| 20 | \| when (v==5)  movetoprevious -> **EMPTY_ROUTE**[8] |
| 21 | \| when (v==1)  waitforloading -> **FULL_ROUTE**[1] |
| 22 | ). |

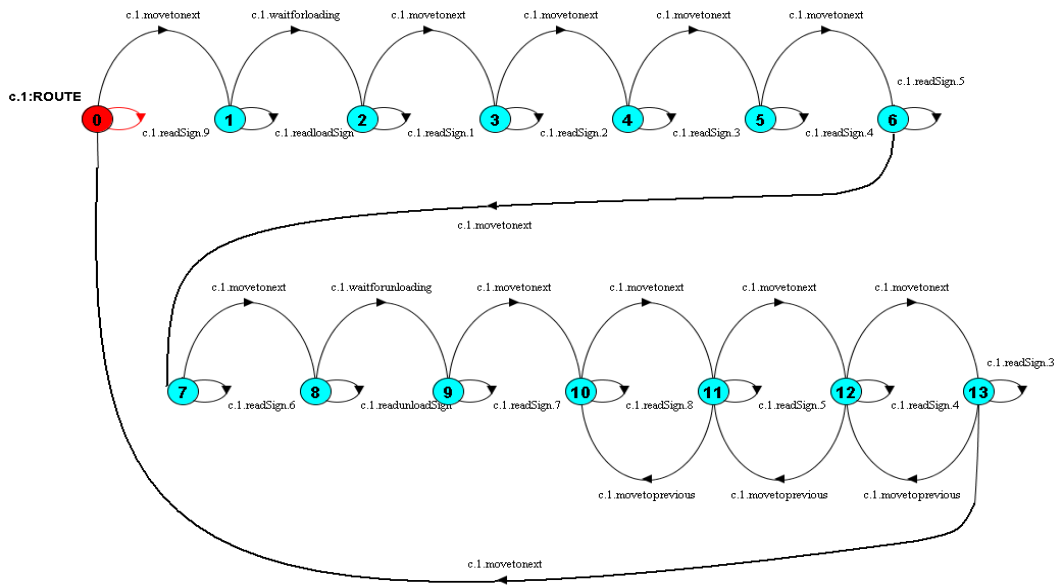

## 7.2. Carrier Agent

The next step is to specify the carrier agents i.e. specify the empty-carrier and full-carrier agents. Here only one carrier agent is taken to represent all the possible states of the system that can arise.

**Fig 11:** LTS specifications of Carrier agent

| | |
|---|---|
| 1 | range R = 1..9 |
| 2 | **CARRIER** = **MOVE_EMPTY**, |
| 3 | **MOVE_EMPTY** = ( |
| 4 | readSign[s:R] -> {movetonext,movetoprevious} -> **MOVE_EMPTY** |
| 5 | \| readloadSign -> waitforloading  -> **MOVE_FULL** |
| 6 | ), |
| 7 | **MOVE_FULL** = ( readSign[s:R] -> movetonext -> **MOVE_FULL** |
| 8 | \| readunloadSign -> waitforunloading -> **MOVE_EMPTY** |
| 9 | ). |

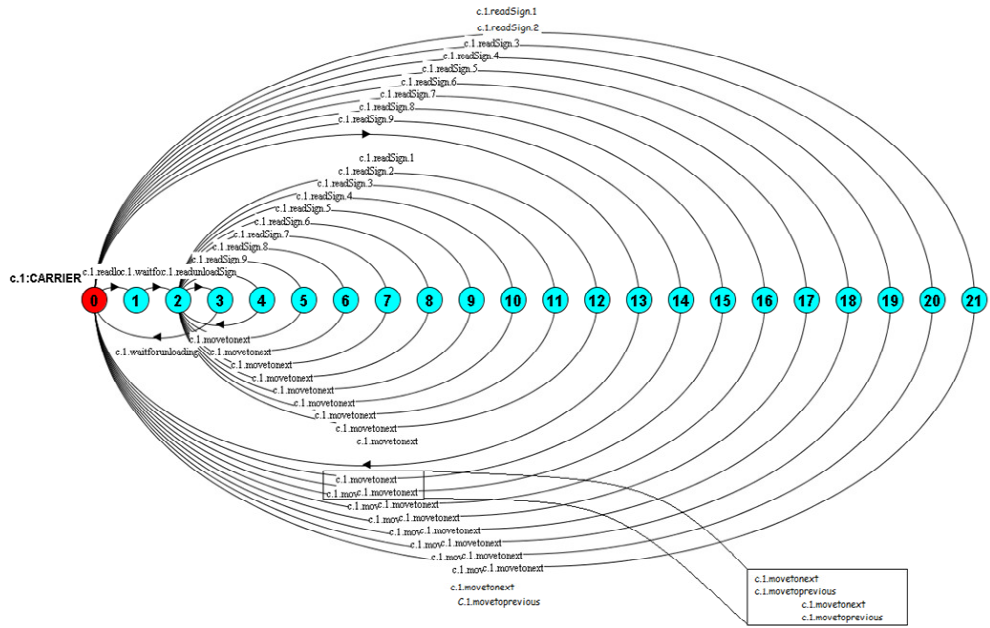

### 7.3. Loader & Un-loader Agents

Loader and un-loader agent loads and un-loads the carrier agents respectively

**Fig 12:** LTS specifications of Loader agent

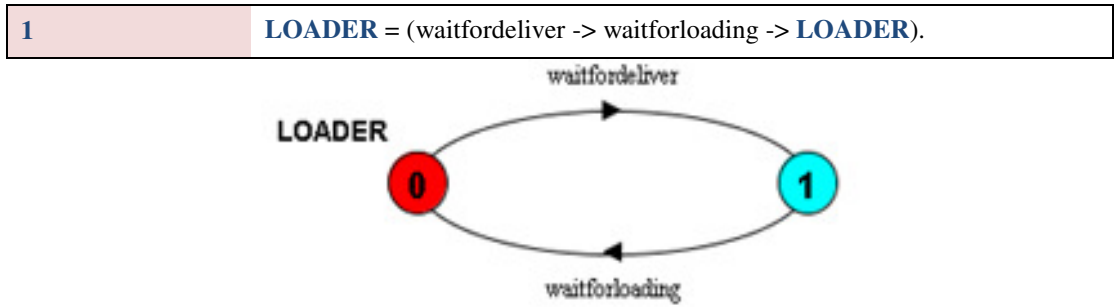

| 1 | **LOADER** = (waitfordeliver -> waitforloading -> **LOADER**). |

**Fig 13:** LTS specifications of Un-Loader agent

| 1 | **UN_LOADER** = (waitforunloading -> waitforreceive -> **UN_LOADER**). |

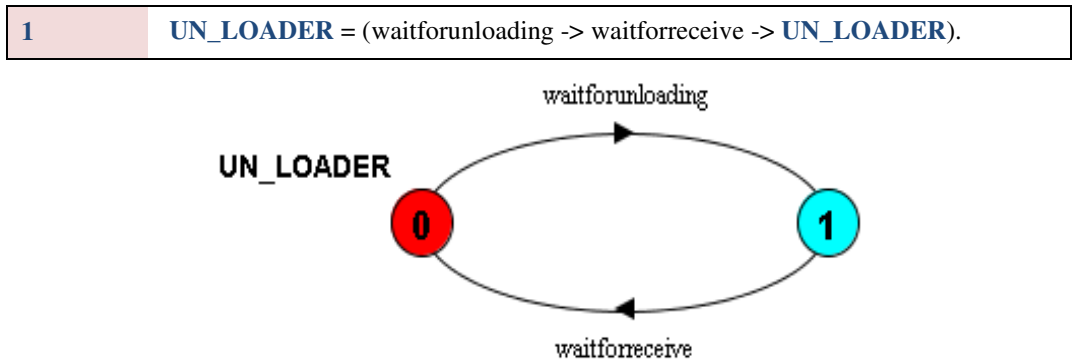

### 7.4. Stock Management

Stock management ensures that the stock at the beginning of the case study at storehouse A is equal to the stock at the end of the case study at storehouse B.

**Fig 14:** LTS specifications of stock management

```
1   const MaxS = 2      /// maximum number of Stock
2   range S = 0..MaxS
3   STOCKFULL_MANAGEMENT = STOCK_FULL[MaxS],
4   STOCK_FULL[st:S] = ( stockCountA[st] -> STOCK_FULL[st]
5   | when(st>0)   decrementStockA -> send -> STOCK_FULL[st-1]
6   | when(st==0)   stockEmptyA -> STOP).
7   STOCKEMPTY_MANAGEMENT = STOCK_EMPTY[0],
8   STOCK_EMPTY[st:S] = ( stockCountB[st] -> STOCK_EMPTY[st]
9   | when(st<MaxS)
10  receive -> incrementStockB ->  STOCK_EMPTY[st+1]
11  | when(st>=MaxS)   stockFullB -> STOP).
12  ||STOCKSYSTEM = (STOCKFULL_MANAGEMENT ||
13  STOCKEMPTY_MANAGEMENT)
14  /{decrementStockA/receive, incrementStockB/send}.
15
16
```

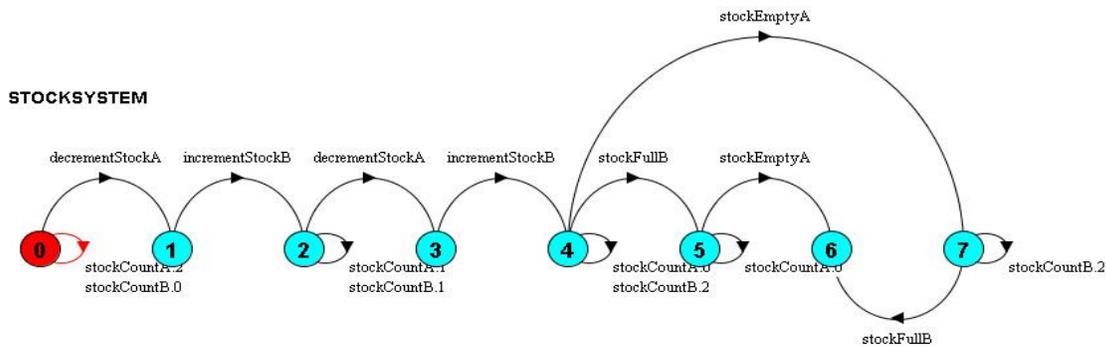

## 7.5. NOLOSS Property

Safety property NOLOSS of Carrier agent infers that there is no loss of stock during the carrier load, unload, and movements between the storehouses. To represent the LTS here with all its states, we have taken a mini-route with only three road partitions. The carrier is loaded and then the carrier is full, there is no loss of stock during the carrier agent's trajectory between storehouse A and B. Safety property specifies every trace that satisfies the property for a particular action alphabet. If the system produces traces that are not accepted by the property automata then a violation is detected during reachability analysis.

**Fig 15:** LTS specifications NOLOSS

```
1    const N=2   // Number of carrier agents
2    const Min=0 // First(Load) road partition
3    const Max=3 // Last(Unload) road partition
4
5    property NOLOSS_Stock = (empty.loaded -> ONTHEWAY[1]),
6
7    ONTHEWAY[part:Min..Max] = (
8    when(part>Min && part<Max)
9    full.moveto[part] -> ONTHEWAY[part+1]
10   | when(part==Max)  full.unloaded -> NOLOSS_Stock).
11   ||NOLOSS = (c[1..N]:NOLOSS_Stock).
12
```

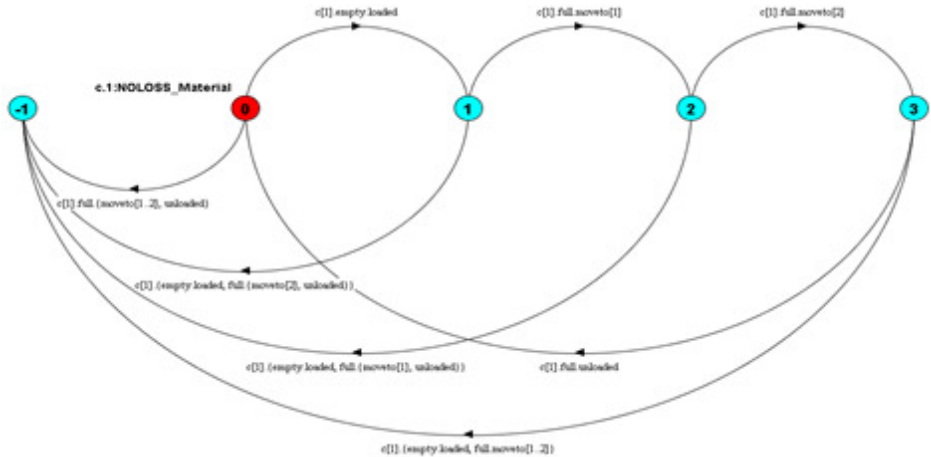

## 8. Future Objectives
The work to be carried out by us in the future is classified into three axes as shown below in fig.16.
1. The automated transformations from one model to another. The development of constructs and tools to generate code automatically from models.
2. Incorporating new versions of the proposed approach. Making improvements in the current approach and proposing new versions which have improved better constructs.

**Fig 16:** Future work: three axes of work to be done

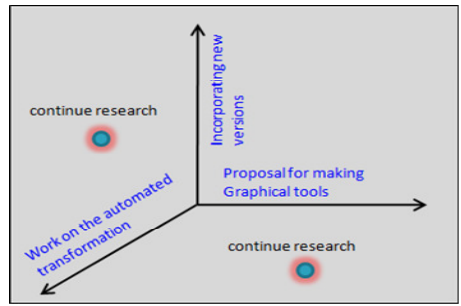

3. Making graphical tools which may lead to an easy drag and drop graphical programming interface for a robotic application development based on π-ADL dot NET language. This programming interface allows novice programmers to program graphically without having knowledge of the underlying rigorous formal methods and languages.

## 9. Concluding Notes
The major contribution is the development of an approach for a multi-agent robotic system that satisfy qualities of correctness i.e. safety and liveness property. An approach based on a combination of methods, formal languages, and techniques is proposed to support the efficient formal description; requirements gathering; formal specification definition; transformation; refinement from abstract to concrete concepts; and verification of multi-agent robotic systems. This approach is exemplified by a case study of a multi-agent robotic system.

     The proposed formal approach phases have key aspects of: organizational abstractions, organizational rules, requirement specifications, role model specifications, protocol definitions, formal requirement verification on the basis of correctness properties, LTS creation, formal static architecture specifications, formal dynamic architecture specifications, and Service Oriented Architecture based simulation implementation. The approach has models based on formal methods and it revolves around

formal verification of correctness properties in each phase from early requirements to the implementation i.e. Gaia method based requirements, Finite state process based finite automata formal verification, π-ADL dot NET language based formal architecture, and services based C# simulation implementation.

The major goal is to facilitate greater assurance to component's correctness. The complete system is specified as a parallel composition of processes and each process synchronizes by means of shared actions. Our system has concurrency, synchronization, correctness, and deadlock issues to be handled and formal model-checking automata-based development methods offer solutions for these issues. Another objective is the use of formal analysis during analysis and design to discover correctness and safety problems early in the development cycle.

## Acknowledgement

We are grateful to Prof. Dr. Muhammad Mukhtar, Vice Chanceller, The Islamia University of Bahawalpur for motivating us for doing research projects, and his support and encouragement for applied research. This work has been possible due to the support of The Department of Computer Science & IT, The Islamia University of Bahawalpur, Pakistan.